\documentclass[journal,comsoc]{IEEEtran}
\usepackage{comment}
\usepackage[T1]{fontenc}

\ifCLASSINFOpdf
\else
\fi
\usepackage{cite}
\usepackage[hidelinks]{hyperref}
\hypersetup{breaklinks=true}
\usepackage{amsmath,amssymb,amsfonts}
\usepackage{algorithmic}
\usepackage{graphicx}
\usepackage{textcomp}
\usepackage{amssymb}
\usepackage{xcolor}
\usepackage[utf8]{inputenc}
\usepackage[T1]{fontenc}
\usepackage{subcaption}
\usepackage{wrapfig}

\usepackage{amsmath}
\usepackage[cmintegrals]{newtxmath}

\usepackage{xcolor}
\usepackage[normalem]{ulem}
\usepackage{cancel}

\usepackage{amsmath}
\usepackage{tabularx}
\begin{document}
\title{Multi-Agent Team Learning in Virtualized Open Radio Access Networks (O-RAN)}
%
%
%


\author{\IEEEauthorblockN{Pedro Enrique Iturria Rivera, Shahram Mollahasani, Melike Erol-Kantarci,~\IEEEmembership{Senior Member,~IEEE}}\\
\IEEEauthorblockA{School of Electrical Engineering and Computer Science \\University of Ottawa} \vspace{-1em}}


\maketitle

\markboth{Submitted to \textbf{\textit{IEEE Wireless Communications Magazine Communications Magazine}}, ON FEBRUARY, 2021}
{Shell \MakeLowercase{\textit{et al.}}: Bare Demo of IEEEtran.cls for IEEE Communications Society Journals}

\begin{abstract}
Starting from the Cloud Radio Access Network (C-RAN), continuing with the virtual Radio Access Network (vRAN) and most recently with Open RAN (O-RAN) initiative, Radio Access Network (RAN) architectures have significantly evolved in the past decade. In the last few years, the wireless industry has witnessed a strong trend towards disaggregated, virtualized and open RANs, with numerous tests and deployments world wide. One unique aspect that motivates this paper is the availability of new opportunities that arise from using machine learning to optimize the RAN in closed-loop, i.e. without human intervention, where the complexity of disaggregation and virtualization makes well-known Self-Organized Networking (SON) solutions inadequate. In our view, Multi-Agent Systems (MASs) with team learning, can play an essential role in the control and coordination of controllers of O-RAN, i.e. near-real-time and non-real-time RAN Intelligent Controller (RIC). In this article, we first present the state-of-the-art research in multi-agent systems and team learning, then we provide an overview of the landscape in RAN disaggregation and virtualization, as well as O-RAN which emphasizes the open interfaces introduced by the O-RAN Alliance. We present a case study for agent placement and the AI feedback required in O-RAN, and finally, we identify challenges and open issues to provide a roadmap for researchers.
\end{abstract}

\begin{IEEEkeywords}
Multi-agent systems, team learning, O-RAN, vRAN.
\end{IEEEkeywords}

\IEEEpeerreviewmaketitle

\section{Introduction}

The demand for mobile connectivity has been undeniably growing over the past decades, including a parallel increase in the demand for better Quality of Service (QoS). On top of that, 5G and the next generations of mobile networks will not only serve smartphone users but also businesses in self-driving cars, healthcare, manufacturing, gaming, marketing, Internet of Things (IoT) and many more. Almost in all generations of mobile networks, resource optimization has been a challenge, yet with 5G, despite new spectrum allocations in the mmWave band, the spectrum is still scarce with respect to increasing demand for wireless connectivity. Moreover, starting from Long-Term Evolution (LTE), the densification trend continues with 5G, at the cost of increased investments, yet, densification is essential due to the limited coverage of mmWave bands. The increasing complexity of mobile networks is reaching the limits of model-based optimization approaches and yielding to data-driven approaches, also known as AI-enabled wireless networks \cite{Elsayed2019}. In this context, the interest for self-optimizing networks that use machine learning is growing \cite{Wang2020a}.   \par

\begin{figure*}
  \includegraphics[width=\textwidth]{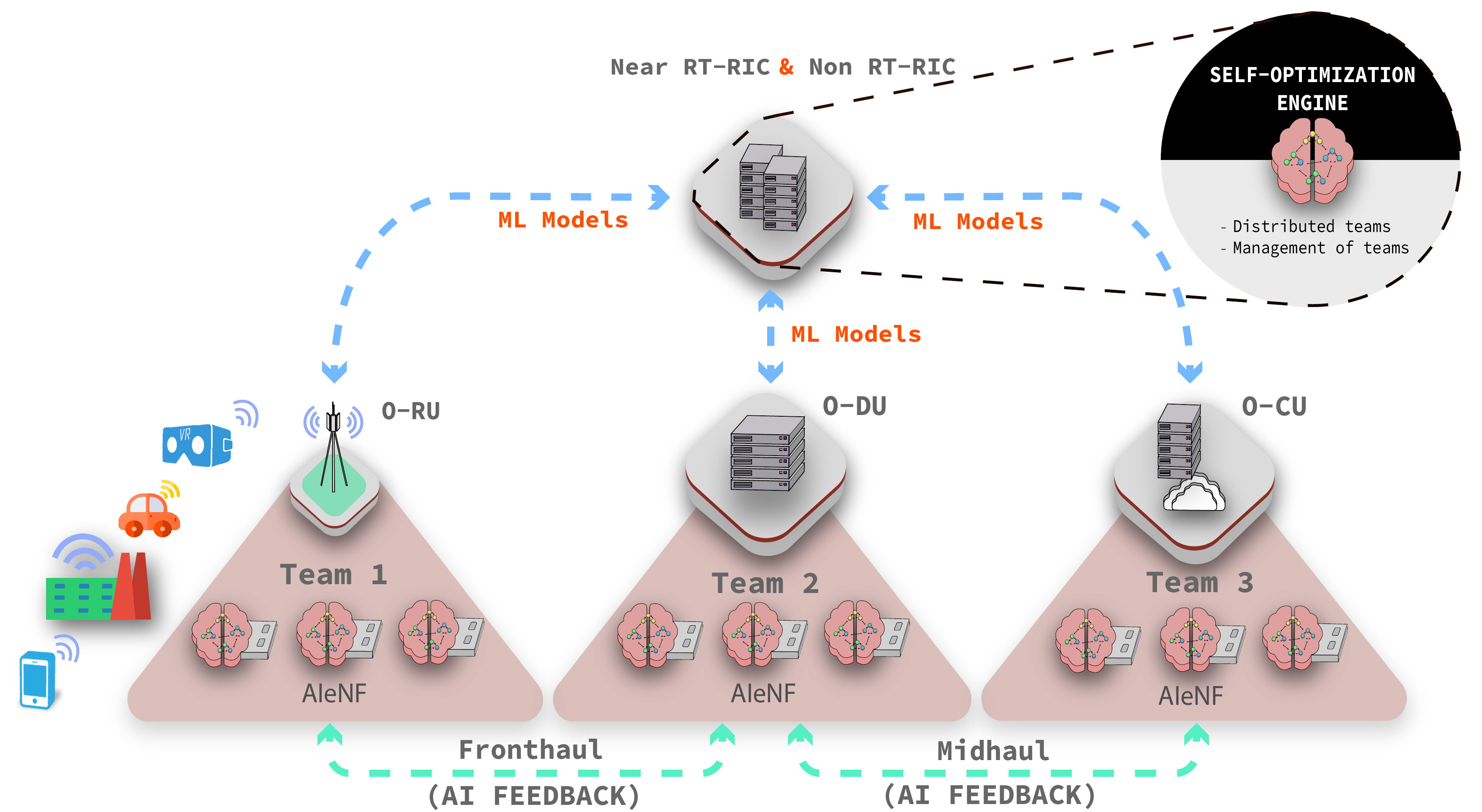}
  \caption{AI-enabled NF agents and team learning in O-RAN. The O-DU (O-RAN Distributed Unit) represents the logical node hosting RLC/MAC/High-PHY layers and the O-RU (O-RAN Radio Unit) is a logical node hosting Low-PHY layer and RF processing. Both cases are based on the 7-2x fronthaul split defined by O-RAN. Finally, O-CU (O-RAN Central Unit) is a logical node hosting RRC (Radio Resource Control), SDAP (Service Data Adaptation Protocol) and PDCP (Packet Data Convergence Protocol) \cite{O-RANWorkingGroup22020}.}
  \label{mas}
\end{figure*}

In the meanwhile, an interesting nexus is emerging from AI-enabled networks and RAN disaggregation and virtualization. In 2010s, Cloud RAN (C-RAN) introduced a disaggregated architecture that allowed grouping Baseband Units (BBUs) into a centralized BBU pool, with the potential of reducing deployment costs in the long term and improving network capacity. Other architectures have continued from the path of C-RAN, such as virtualized RAN (vRAN) which adds the concept of hardware functions being softwarized. RAN disaggregation and virtualization offer many advantages over traditional RAN solutions, such as reduced network deployment time, improved energy efficiency and enhanced mobility support. Most recently, Open Radio Access Network (O-RAN) Alliance has brought together those prior efforts around a new RAN architecture with open interfaces, aiming to replace some of the interfaces that have become proprietary over time due to vendor specific implementations of standards. Open interfaces are crucial for equipment from multiple vendors to be stacked together and for the selection of best-of-breed solutions. 

The timing of O-RAN coincides with new advances in Software Defined Networking (SDN), Network Function Virtualization (NFV), dynamic function splitting, high capacity data centers, cloud and edge computing, all of which have set the stage for O-RAN \cite{Chih-Lin2020}. Meanwhile the increased complexity and the flexible architecture of the O-RAN specifications call for the use of machine learning techniques more than ever.  \par

As of 2021, there are a handful of implementations of open RAN around the globe. For instance, in Japan, Rakuten has deployed distributed data centres hosting both Distributed Unit (DU) and Central Unit (CU) functions. The Spanish operator Telefónica has invested in the integration of its open RAN solution with its multi-access edge computing (MEC) solution. 
In addition, some operators such as Sprint and T-Mobile in the US and Etisalat in the middle-east are in the testing phase of open RAN. There are many other emerging vendors, system integrators and operators who are experimenting with open RAN, in addition to the above mentioned players in the market. Some of these implementations follow O-RAN specifications.   

The multitude of layers, functions, splits, and the consequent complexity in O-RAN, position machine learning techniques that involve multiple agents and team learning, as potential solutions. In recent years, MASs have been used to address complex problems in robotics and other self-driving systems therefore they can be promising alternatives for the self-driving (or self-optimizing) capability of O-RAN. To the best of our knowledge, this paper is the first to elaborate on multi-agent team learning and its use in O-RAN. Fig. \ref{mas} illustrates AI-enabled Network Function (AIeNF) agents being deployed within the O-RAN architecture where they are organized as teams by the O-RAN controller. 
The purpose of this paper is not to provide a survey of MAS or an introduction to O-RAN specifications. In the literature, there are several comprehensive surveys on MAS. For instance, in \cite{Dorri2018}, the authors provide an extensive discussion of all aspects of MASs, starting from definitions, features, applications, challenges, and communications to evaluation. 
This paper focuses on the application of multi-agent team learning in the O-RAN environment. We first begin with an introduction on MAS and team learning in MAS, and then continue with presenting background on recent advances in RAN architectures. We then show a case study for agent placement in O-RAN from AI feedback perspective and discuss the range constraints. Finally, we provide a detailed discussion on the challenges of multi-agent team learning in O-RAN, and identify open issues in the field.


\section{Background on Multi-Agent Systems} 
A MAS is a group of autonomous and intelligent agents that act in an environment in order to accomplish a common goal or individual goals. There can be several types of MAS, such as homogeneous/heterogeneous, communicating/non-communicating and cooperative(collaborative)/competitive.

\textbf{Homogeneous}: The multi-agent system consists of agents with a similar internal structure, which means that all agents have the same local goals, capabilities, actions, and inference models. In the homogeneous architecture, the main difference among agents is based on the place where their actions are applied over the environment.

\textbf{Heterogeneous}: In the heterogeneous MAS, agents may have different goals, capabilities, actions, and inference models.  

From the O-RAN perspective, homogeneous agents would be instances of the same network function (NF) (e.g. resource allocation), while different network functions (e.g. resource allocation and mobility management) working together in the RAN stack would be heterogeneous agents.   


\textbf{Communicating/non-communicating:} A group of agents can be designed to communicate with each other or not. When there is no communication, agents act independently, and they don't use feedback from other agents however they may be working in the same environment hence receive indirect feedback on the actions of the other agents. In the O-RAN architecture, the way the agents communicate will determine the amount of bandwidth consumed on the interfaces. Furthermore, in a multi-vendor environment some agents might not communicate with the others.


 

In a MAS, agents can be cooperative, competitive, or mixed. In general, if they are cooperative, agents communicate with each other. For the competitive case, they may or may not communicate, or they may share partial information. However, their  behavior on cooperation or competition is different than whether they communicate or not. A group of agents may communicate and can still be competing.

\textbf{Cooperative}: In a cooperative setting, the agents need to take collaborative actions to achieve a shared goal, or in other words, agents must jointly optimize a single reward signal.

\textbf{Competitive}: In a competitive setting, each agent tries to maximize one’s own reward under the worst-case assumption meanwhile the other agents always try to minimize the reward of others.

It is also possible to have groups with mixed behavior agents where some are cooperative and some are competitive. In the O-RAN architecture, this kind of behavioral differences may be needed for various use cases, where in a chain of NFs (such as in Service Function Chains (SFCs)), cooperation would be preferred, while between different slices competition might be observed.   


Although many machine learning techniques have been considered for MAS, team learning could have particular significance in the future, since it can have applications in O-RAN by organizing the intelligent NFs as teams of teams, in hierarchies. In the next section, we summarize the team learning techniques in MAS. 

\subsection{Team Learning in MAS}
In team learning, teams of agents are usually assumed to be distributed, cooperative, and completely share information on observations. However, it is also possible to consider mixed behavior or competitive MAS with complete or partial observability (i.e. partial information sharing). Various ways of organizing intelligence as well as use cases in O-RAN would require its own team model.

Team learning has been studied and applied for a wide range of applications outside the wireless research. From multi-robot teams (e.g., robot soccer), multi-player games (e.g., Dota, StarCraft), predator-prey pursuit and capture problems, to search and rescue missions, team learning has been used to address complex problems where collaboration or competition among agents is present. Team learning has not been used in wireless systems, however in the future wireless networks, there will be a need for organization of intelligence in teams. Therefore, we explore applications of team learning in different disciplines and discuss how they can align with wireless systems.

Recent applications of team learning in the scope of multi-robot teams use reinforcement and deep learning techniques. For instance, in  \cite{Wang2020} the authors propose a novel approach for coordination among a team of robots using their relative perspective-based image. 
The goal is to deal with resource competition and static and dynamic obstacle avoidance problems among the team members. 

Search and rescue mission is another application type where team learning has been of great interest due to its practicality in hostile environments. In \cite{Gunn2015}, the authors study the behavior of heterogeneous robots in damaged structures. A framework on how team maintenance and task management among teams is described in this work. 

Finally, an interesting work related to team learning in urban network traffic congestion is presented in \cite{Pan2020}. The authors proposed a Dynamic Traffic Assignment (DTA) algorithm based on collaborative decentralized heterogeneous reinforcement learning approach to mitigate the effects of the randomness of urban traffic scenarios. In this setup, two agent groups are defined as advisers and deciders. The deciders assign the flows of traffic network meanwhile the advisers communicate support information exclusively with the deciders. 
In particular, adviser-decider type of team learning can find use in O-RAN considering the different time granularities of non-Real-Time RIC (non-RT RIC) and near-Real-Time RIC (near-RT RIC). For instance, a team of agents at non-RT RIC can advice a team of near-RT RIC agents on long term policies.
In the next section, to complete the picture, we first provide background on RAN disaggregation, virtualization and O-RAN, and then overview MAS applications in this domain.

\section{Disaggregated, Virtualized RAN and O-RAN}

Earlier RAN solutions offered an architecture where BBUs and RUs were co-located. This brought limitations in terms of not being able to pool BBU resources. Therefore the following phase of RAN architectures considered BBU resources that are pooled close to the radios but not co-located, where geographical proximity is necessary due to latency limitations. The pool of BBUs is called Distributed Unit (DU), and the radios constitute the Radio Unit (RU). Within O-RAN specifications, another level of processors is also defined which is called as O-RAN Central Unit (O-CU). 

The most appealing reason behind RAN disaggregation was to reduce costs and bring more versatility to the technological market. An earlier version of RAN disaggregation was seen in C-RAN where some hardware functions are implemented as software functions and BBU functionality is collected at the centralized cloud. C-RAN offers improvements in network capacity, handling cooperative processing, mobility, coverage,  energy-efficient network operation and reduced deployment costs \cite{Checko2015}. 

On the evolution path of RAN architecture, the most recent development comes with O-RAN, in which interfaces between O-RU and O-DU and between O-DU and O-CU are based on open specifications, meaning inter-operability between vendor products and the possibility to select the best-of-breed technology by Mobile Network Operators (MNOs). In addition, O-RAN embraces intelligence in every layer of its architecture and aims to leverage new machine learning-based technologies. 
A recent survey on open virtualized networks \cite{Bonati2020}, gives a comprehensive overview of the state-of-the-art in modern RANs.

In the next section, we summarize the efforts in the intersection of AI and O-RAN, as well as the use of MAS in state-of-the-art wireless networks.

\subsection{The Role of Intelligence in O-RAN}
\begin{figure*}
	\includegraphics[width=\linewidth]{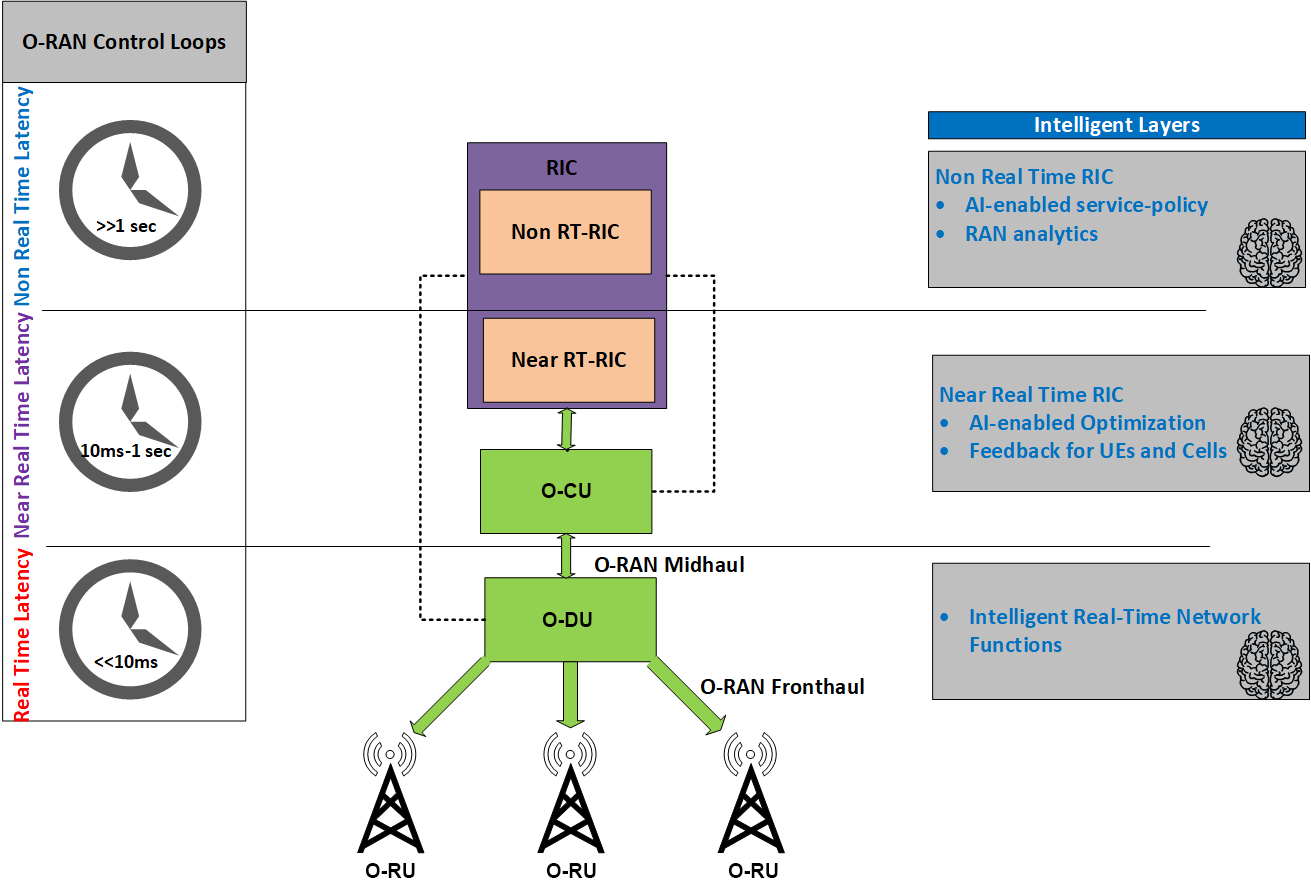} 
	\caption{O-RAN closed-loop latency requirements and the intelligence at different layers of O-RAN.} 
	\label{o-ran}
\end{figure*}

O-RAN defines certain guidelines to employ AI in its architecture. Offline and online learning is expected to coexist with a modular design as best practice to follow. This will enable service providers to decide the location of intelligence in the network functions according to their best interests. As recommendation, applications are expected to fall in certain control loops according the time budget needed for such applications (see Fig. \ref{o-ran}). Furthermore, open source solutions, such as Acumos AI, emerge as a potential development platform for ML models. 
Several AI/ML O-RAN use cases are already identified, such as QoE (Quality of Experience) optimization, traffic steering and V2X handover management \cite{O-RANWorkingGroup22020}. It is important to highlight that although some ML applications are described in O-RAN reports, the huge potential of applying multi-agent team learning in O-RAN are not covered in any of the use cases.  

On the other hand, MAS has been studied in the context of wireless networks in a few studies. In \cite{Yan2018}, the authors address resource scheduling in a multi-RAT environment by proposing a Smart Aggregated RAT Access (SARA) strategy based on multi-agent reinforcement learning to maximize the average system throughput while satisfying the UEs' QoS preferences. The proposed algorithm consists of a model-free multi-agent reinforcement learning to solve a Semi-Markov Decision Process (SMDP) based on Hierarchical Decision Framework (HDF). Multi-RAT solutions are particularly important for O-RAN since some ideas can be employed in multi-vendor multi-RAT systems.

Most of the prior work on machine learning in wireless networks consider multiple uncoordinated agents working within the same environment, driven by the same rewards or goals. However, when multiple agents interact with the environment independently, thus, changing the environment of each other, or when they have different goals, the problem cannot be simplified to deploying independent agents. Therefore, using teams of agents in O-RAN emerges as a promising approach. Yet, the communication overhead among agents need to be considered. As a first step towards team learning within O-RAN, in the next section, we show the tradeoff between AI feedback due to agent placement and the fronthaul range.

\section{A Case Study on Team Agent Placement and O-RAN Midhaul Range}
As the O-RAN architecture attempts to embed intelligence into multiple layers and functions, it is natural to consider that intelligence will be distributed. Accordingly, this provides the possibility of invoking intelligent NFs at different computational resources, by dynamic functional splits, to either access higher processing power to optimize more complex problems (e.g. at the O-CU) or apply low-latency actions for low latency services (e.g. at the O-DU). When agents in a team are invoked at different places, they will require feedback from the other agents on the observed environment. We call this as AI payload and investigate the trade off between AI payload and the physical midhaul range between compute resources in O-RAN.  

As it is shown in Fig.\ref{o-ran}, each O-RAN component needs to abide by a certain delay bound that is tolerable within its control loop. This is closely related to the bandwidth and the range between O-CU, O-DU and O-RU. Since, most intelligent NFs will be distributed to O-CU and O-DU, midhaul range is of particular interest. To determine the midhaul range for a certain amount of AI payload, maximum tolerable delay needs to be considered. The separation of network functions between the compute resources is referred to as the functional split. In 3GPP eight functional splits have been defined and the implementation of splits has been of great interest given the fact that it is possible to balance the trade off between the fronthaul/midhaul traffic and the latency \cite{Matoussi2020}. Here, we consider dynamic functional split as intelligent NF agents being placed at either O-CU or O-DU, and we study the limit on midhaul range based on  AI payload that needs to be exchanged between those agents. For this, the overall delay of the closed-loop control can be represented by the round trip time (RTT) of AI feedback, which includes processing, transmission and propagation delay over the midhaul. The maximum allowable range of the midhaul is proportional to the propagation delay of the physical link. As shown in Table \ref{orabtab}, each functional split option has its own delay budget and bandwidth requirement. Hence, the maximum allowed one-way latency presented in the table becomes the limiting factor for the maximum allowable range \cite{ITU-TRecommendationGSuppl.662018}. Since machine learning algorithms will be using the AI payload as the feedback that is delivered on the midhaul, it is important to consider these latency budgets and ranges accordingly. In \cite{Pamuklu2020} dynamic functional splits, using reinforcement learning, is considered in O-RAN to maximize the use of renewable energy. Here, we show the relation between the AI payload and the range limitation.   

\begin{table}[]

\caption{Bandwidth, latency and range limits for functional splits. Bandwidth and latency figures are taken from \cite{ITU-TRecommendationGSuppl.662018}.}
\begin{tabular}{|c|c|c|c|}
\hline
Split Option & \begin{tabular}[c]{@{}c@{}}Minimum required \\ \\ bandwidth\end{tabular} & \begin{tabular}[c]{@{}c@{}}Maximum allowed\\ \\ one-way Latency\end{tabular} & \begin{tabular}[c]{@{}c@{}}midhaul\\ range\\ (km)\end{tabular} \\ \hline
Option 1     & 3 Gbps                                                                   & 10 ms                                                                        & 400                                                              \\ \hline
Option 2     & 3 Gbps                                                                   & 1 ms                                                                         & 40                                                               \\ \hline
Option 3     & 3 Gbps                                                                   & 1.5 ms                                                                       & 60                                                               \\ \hline
Option 4     & 3 Gbps                                                                   & 100 $\mu$s                                                                   & \textless{}1                                                     \\ \hline
Option 5     & 3 Gbps                                                                   & 100 $\mu$s                                                                   & \textless{}1                                                     \\ \hline
Option 6     & 4.1 Gbps                                                                 & 250 $\mu$s                                                                   & 2.7                                                              \\ \hline
Option 7a    & 10.1 Gbps                                                                & 250 $\mu$s                                                                   & 7                                                                \\ \hline
Option 7b    & 37.8 Gbps                                                                & 250 $\mu$s                                                                   & 9.2                                                              \\ \hline
Option 7c    & 10.1 Gbps                                                                & 250 $\mu$s                                                                   & 7                                                                \\ \hline
Option 8     & 157.3 Gbps                                                               & 250 $\mu$s                                                                   & 10                                                               \\ \hline

\end{tabular}
\label{orabtab}
\end{table}

\begin{figure}
	\includegraphics[width=\linewidth]{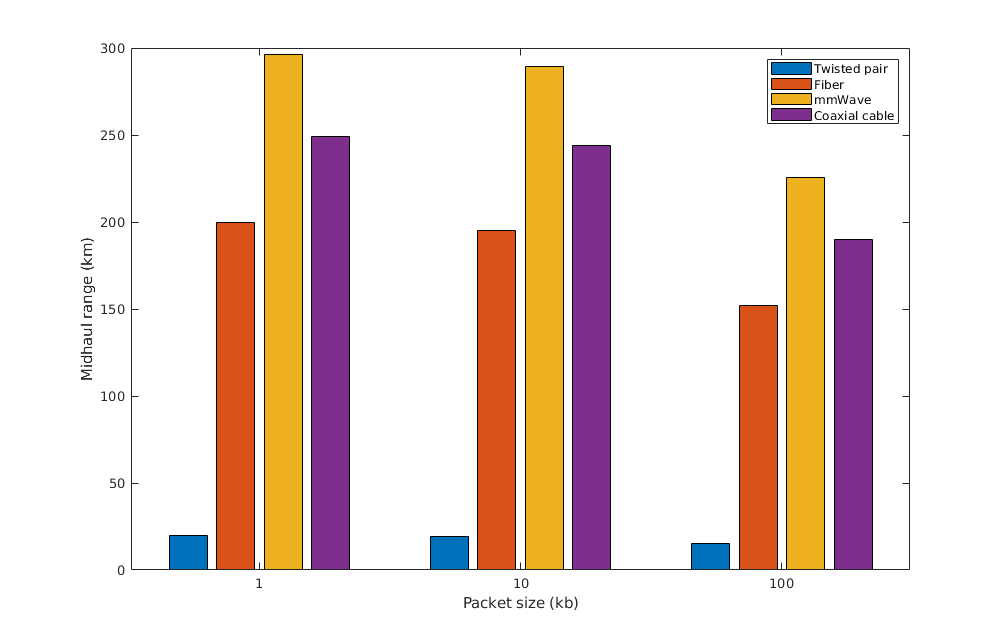} 
	\caption{Maximum midhaul range for varying AI feedback payload.} 
	\label{technologies}
	
\end{figure}

Note that, the type of medium used for midhaul is a primary factor that affects the range. For instance,  the propagation delay in fiber optic is around  5 $\mu s/km$. This value can be reduced to 4 $\mu s/km$ and 3.34 $\mu s/km$ if coaxial cables and mmWave technologies are employed, respectively. At the same time, the propagation delay in Ethernet is 50 $\mu s/km$, which is much higher with respect to fiber optic, mmWave, and coaxial cables. As a case study, we focus on midhaul and in  Fig.~\ref{technologies}, we provide the maximum range for several technologies under varying packet sizes (AI payload). Although currently, in mmWave the communication range is small, we assume the communication range can be extended by employing multiple hubs equipped with these technologies. It is important to note that, as the payload of AI increases (i.e. larger packets), the O-DU and O-CU needs be located closer. 

While designing team learning algorithms or other MAS techniques, the amount of communication between agents becomes a critical issue.  In addition, although implementing intelligent agents at O-CU can provide broader perspective of intelligence due to the accessibility of other agents' observations and access to more processing power, the location of agents and the amount of data that needs to be transmitted between O-CU and O-DU can negatively impact the network performance. Therefore, when invoking teams of agents at different layers of O-RAN, bandwidth and range of the midhaul need to be taken into consideration. 
In the next section, we further elaborate on the other open issues, as well as the opportunities of multi agent team learning in O-RAN.

\section{Open Issues and Future Directions for Multi Agent Team Learning in O-RAN}
Despite many potential benefits of machine learning in O-RAN, it is hard to deny the numerous challenges that need to be addressed before AI or ML can take over RAN functions. In this section, we identify the challenges, open issues and opportunities in multi-agent team learning for O-RAN. Note that, some challenges are already fundamental issues in learning in the MAS environment  while others are only specific to O-RAN.  
\begin{itemize}
\item \textbf{Convergence:}
Similar to all other machine learning-based techniques, team learning for the O-RAN should converge. Divergent behavior will cause instability. However, convergence guarantee in decentralized learning for stochastic teams is known to be a challenging problem \cite{Yongacoglu2019}. In addition, for the RAN environment convergence should be fast. In this case, fast boot strapping techniques or offline training needs to be considered.

\item \textbf{Scalability:} Dimensionality issues have been a recurrent issue when the number of agents (so as the states and actions) in a MAS tends to increase. As more NFs become intelligent agents in O-RAN, the scalability of learning, inter-agent communication and environment feedback need to be addressed.

\item \textbf{Lack of full observability:} As intelligent NFs act in an environment, they will be simultaneously modifying the environment of other agents. Therefore, to take an optimal action, each agent will need to predict other agents’ actions unless all agents can fully observe the environment which unlikely in O-RAN. Hence, decisions need to be made based on the agents’ partial observations, which can result in a sub-optimal solution. ML models that are robust under partial observability need to be explored.
\item \textbf{Information sharing:} It is essential to decide how much of the available local information should be shared among agents for enhancing the modeling of an environment. As addressed in the previous section, this should be jointly considered with fronthaul and midhaul technologies and the functional split decisions. 
\item \textbf{Selecting the optimal team size:}  Choosing the optimal team size can affect learning and system performance. Although a larger team can provide wider visibility over the environment and access more relevant information, the incorporation and learning experience of each agent can be affected. Meanwhile,  one can obtain faster learning within a smaller team size, but due to the limited system view, a sub-optimal performance may be achieved. Optimal team organization will be fundamental to O-RAN. 

\item \textbf{Goal selection:} In a RAN environment, the agents may reach some decisions where the network performance is degraded, even though the individual agents intend to maximize the performance. The goal of team learning should be minimizing conflicts and focusing on overall performance enhancement.

\item \textbf{Impact of delayed feedback and jitter:}
Most MAS studies consider that the feedback from the environment is immediate, and if agents are communicating, their feedback is instant. However, in disaggregated RANs, feedback can be intentionally or unintentionally delayed. Delayed feedback may cause agents to interact with diverged versions of the environment, and lead to degraded RAN performance. 

\item \textbf{Security and trust:} Most MAS rely on the truthfulness of information shared among agents. Although there are studies on uncertainty or partial observability, intentional wrong reporting and adversarial behavior should also be considered. Considering the high stakes of an attack in the wireless infrastructure, this concern cannot be left unaddressed \end{itemize}

\section{Conclusion}
In this paper, we motivate the use of multi-agent team learning as a way of organizing intelligence in O-RAN. We first gave an overview of MAS and team learning, and then we provided a short summary of RAN evolution from C-RAN to vRAN to O-RAN. We presented a case study on the requirements of midhaul range when intelligence is embedded at different layers of O-RAN. Finally, we provided a detailed discussion on challenges, open issues and future directions for multi-agent team learning in O-RAN.


\section{Acknowledgment }
This research was supported by the Natural Sciences and Engineering Research Council of Canada, Canada Research Chairs program.

\bibliography{references.bib}{}
\bibliographystyle{IEEEtran}

\begin{IEEEbiography}[{\includegraphics[width=1in,height=1.25in,clip,keepaspectratio]{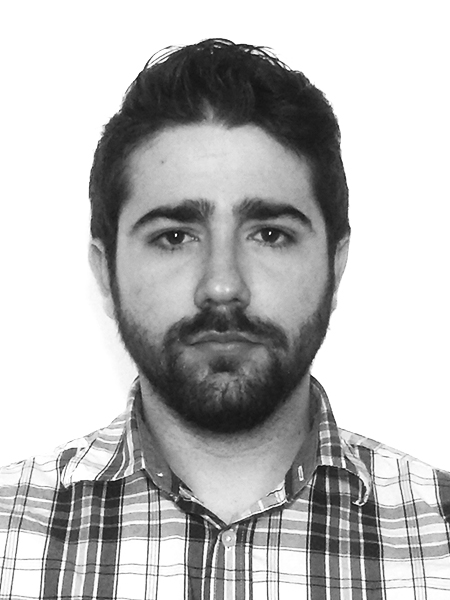}}]{Pedro Enrique Iturria Rivera}
is a PhD student at the School of Electrical Engineering and Computer Science at the University of Ottawa. He is a member of the Networked Systems and Communications Research Laboratory (NETCORE). His research interests include distributed multi-agents for wireless networks, smart grid communications, data privacy, data science and color science.
\end{IEEEbiography}

\begin{IEEEbiography}[{\includegraphics[width=1in,height=1.25in,clip,keepaspectratio]{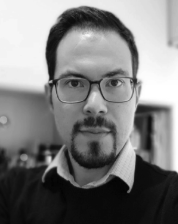}}]{Shahram Mollahasani}
received his PhD degree from the Middle East Technical University in 2019. Since 2019, he is a Post-Doctoral Fellow at the University of Ottawa and a member of the Networked Systems and Communications Research Laboratory (NETCORE). His research interests include wireless communications, green networks, and 5G and beyond mobile networks. He is a member of the IEEE.   
\end{IEEEbiography}

\begin{IEEEbiography}[{\includegraphics[width=1in,height=1.25in,clip,keepaspectratio]{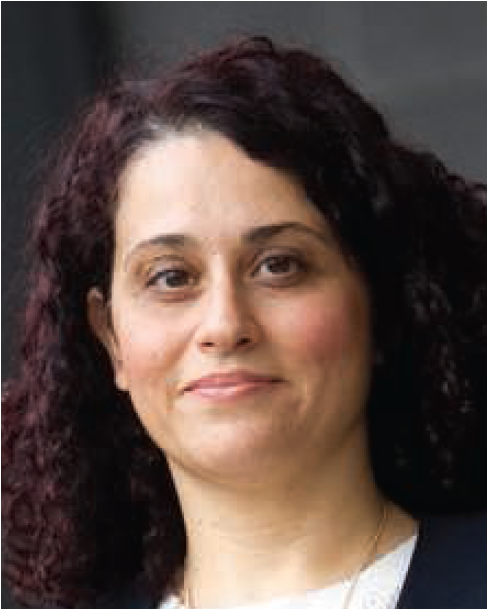}}]{Melike Erol-Kantarci}
is Canada Research Chair in AI-enabled Next-Generation Wireless Networks and Associate Professor at the School of Electrical Engineering and Computer Science at the University of Ottawa. She is the founding director of the Networked Systems and Communications Research (NETCORE) laboratory. She has received several best paper awards including the IEEE Communication Society Best Tutorial Paper Award in 2017. Dr. Erol-Kantarci is the co-editor of three books on smart grids, smart cities and intelligent transportation. She is on the editorial board of the IEEE Transactions on Cognitive Communications and Networking, IEEE Internet of Things Journal, IEEE Communications Letters, IEEE Networking Letters, IEEE Vehicular Technology Magazine and IEEE Access. She has acted as the general chair and technical program chair for many international conferences and workshops. Her main research interests are AI-enabled wireless networks, 5G and 6G wireless communications, smart grid, electric vehicles, Internet of things and wireless sensor networks. She is a senior member of the IEEE and ACM.
\end{IEEEbiography}

\end{document}